\documentclass[]{aa} 

\def\deg{$^\circ$}
\def\arcsec{"}

\def\aap{A\&A}

\def\apjl{ApJL}

\def\kms{${\rm km}\cdot{\rm s}^{-1}$\,}

\def\msun{M$_{\odot}$\,}

\usepackage{graphicx}
\usepackage{txfonts}
\usepackage{natbib}

\begin{document}

\title{The expanding dusty bipolar nebula around the nova V1280 Sco\thanks{Based on observations made with ESO telescopes
at the Silla Paranal Observatory under programme IDs 283.D-5020, 385.D-0076, 087.D-0074}}

\authorrunning{Chesneau, Lagadec et al.}
\titlerunning{A bipolar dusty nebula around the nova V1280 Sco}

   \author{O. Chesneau
          \inst{1}
       \and E. Lagadec \inst{2}
              \and M. Otulakowska-Hypka \inst{3}
              \and D.P.K. Banerjee\inst{4}
              \and C.E.~Woodward\inst{5}
              \and E.~Harvey\inst{1}
               \and A.~Spang\inst{1}
               \and P.~Kervella\inst{6}
               \and F.\,Millour\inst{1}
               \and N.\,Nardetto\inst{1}
               \and N.M.\,Ashok\inst{4}
               \and M.J.\,Barlow\inst{7}
               \and M.\,Bode\inst{8}
               \and A.\,Evans\inst{9}
               \and D.K.\,Lynch\inst{10}
               \and T.J.\,O'Brien\inst{11}
               \and R.J.\,Rudy\inst{10}
               \and R.W.\,Russel\inst{10}
    				} 
          

   \offprints{O. Chesneau}

   \institute{Laboratoire Lagrange, UMR7293, Univ. Nice Sophia-Antipolis, CNRS, Observatoire de la C\^ote d'Azur, 06300 Nice, France\\
              \email{Olivier.Chesneau@oca.eu}%
                \and
European Southern Observatory, Karl-Schwarzschild-Strasse 2 D-85748 Garching bei M\"unchen, Germany
   \and
N. Copernicus Astronomical Center, Polish Academy of Sciences, Bartycka 18, 00-716 Warsaw, Poland
   \and
Physical Research Laboratory, Navrangpura, Ahmedabad, Gujarat, India
\and
Minnesota Institute for Astrophysics, University of Minnesota, Minneapolis, MN 55455, USA
\and
LESIA, Observatoire de Paris, CNRS UMR 8109, UPMC, Universit\'e Paris Diderot, 5 place J. Janssen, F-92195 Meudon, France
\and
Department of Physics and Astronomy, University College London, Gower Street, London WC1E 6BT
\and
Astrophysics Research Institute, Liverpool John Moores University, Birkenhead, CH41 1LD, UK
\and
Astrophysics Group, Keele University, Keele, Staffordshire ST5 5BG
\and
Space Science Applications Laboratory, The Aerospace Corporation, M2/266, P.O. Box 92957, Los Angeles, CA 90009
\and
Jodrell Bank Cent. for Astrophysics, School of Physics and Astronomy, Univ. of Manchester, Oxford Road, Manchester M13 9PL
}
   \date{Received ;accepted }

 
  \abstract
   {The fast temporal evolution of the ejecta morphology of novae can be considered 
   as an important test bench for studying the shaping of many kinds of nebulae. \object{V1280~Sco} is one of the 
   slowest dust-forming nova ever historically observed that has experienced a particularly 
   long common-envelope phase.}
   {We performed multi-epoch high-spatial resolution observations of the circumstellar 
   dusty environment of \object{V1280\,Sco} to investigate the level of asymmetry of the ejecta.}
   {We observed \object{V1280\,Sco} in 2009, 2010 and 2011 (from t=877 days after discovery until t=1664\,d) 
   using unprecedented high angular resolution techniques. We used the NACO/VLT 
   adaptive optics system in the J, H and K bands, together with contemporaneous VISIR/VLT
   mid-IR imaging that resolved the dust envelope of \object{V1280~Sco}, and SINFONI/VLT observations secured in 2011. }
   {We report the discovery of a dusty hourglass-shaped bipolar nebula. The apparent 
   size of the nebula increased from 0.30\arcsec x 0.17\arcsec\ in July 2009 to 0.64\arcsec x 0.42\arcsec\ in
   July 2011. The aspect ratio suggests that the source is seen at high inclination. The central
   source shines efficiently in the K band and represents more than 56$\pm$5\% of the total flux in 2009, 
   and 87$\pm$6\% in 2011. A mean expansion rate of 0.39$\pm$0.03 milliarcsec per day is inferred from 
   the VISIR observations in the direction of the major axis, which represents a projected upper limit.
   Assuming that the dust shell expands in that direction as fast as the low-excitation slow ejecta detected
   in spectroscopy, this yields a lower limit distance to V1280 Sco of $\sim 1$~kpc; however, the systematic errors remain large due to the complex shape and velocity field of the dusty ejecta.  The dust seems to reside essentially in the polar caps and no infrared flux is detected in the equatorial regions in the latest dataset. This may imply that the mass-loss was dominantly polar.
}
   {\object{V1280~Sco} is an excellent test case for studying the temporal evolution of dusty bipolar ejecta. As the nebula expands, observations will be easier and we advocate a yearly
   monitoring of the source using high angular resolution techniques.}

                
   \maketitle
%

\section{Introduction}

White dwarfs (WDs) are the end products of the evolution of low- and intermediate-mass stars,
with initial masses between 0.8 and 8 M$_{\sun}$. If a white dwarf is in a  close binary system,
it can accrete hydrogen from its close companion's atmosphere, which then triggers the ignition of 
runaway nuclear reaction. This leads to a bright outburst, called a classical nova.

The classical nova \object{V1280\,Sco} was discovered in outburst by \citet{2007IAUC.8803....1Y} on 2007 
February 04.86 (JD = 2 454 136.85), about 12 
days before reaching its maximum in visual light (m$_{\rm V}\sim$4) and formed dust only two weeks after
reaching maximum in visible \citep{2008MNRAS.391.1874D}. This nova experienced one of the slowest
evolutions historically reported and the nebular phase was entered about 50 months (t$\sim$1600d) after outburst \citep{2012arXiv1203.6725N}. This is a
clear indication that the mass of the erupting WD is low (i.e. $\sim$0.6\msun\ or even smaller).

Extensive monitoring of the event was performed using the Very Large Telescope
Interferometer (VLTI) during the first four months (until t=145d) of the event  \citep{2008A&A...487..223C}.
Owing to the sparse $uv$ coverage, an interpretation involving a spherical dusty shell was
developed, providing estimates of the dust formation rate. Based on an estimate of the total dust mass
formed during  the 250 days after outburst, \citet{2008A&A...487..223C} suggested that a mass as large
as 10$^{-4}$ \msun was ejected. \citet{2012arXiv1203.6725N} provided an independent
estimate from neutral oxygen lines that confirmed the high mass ejected, an additional 
argument for a low-mass WD progenitor, known to statistically eject more mass that fast, massive ones.
The interferometric
observations also provided a consistent picture of the expansion rate in the plane of the
sky estimated at 0.35\,mas/d. Radial velocities were published by
\citet{2008MNRAS.391.1874D} in the near-IR domain, and by \citet{2010PASJ...62L...5S} 
in the visible. \citet{2010PASJ...62L...5S} reported multiple high-velocity narrow components
in the NaI D line from 650 to 900 km.s$^{-1}$. They interpreted this as being
due to the formation of  clumps, 
as a consequence of a strong shock between the slow, dust-forming ejecta, and a  
fast wind ($\sim$ 2000 km.s$^{-1}$) generated during the second brightening of the
source \citep[t=110d]{2008A&A...487..223C}. P-Cygni lines were systematically observed even
four years after the outburst, which is indicative of a sustained and significant mass loss \citep{2011ApJS..197...31S}. The typical P-Cygni type velocities reached
500 km.s$^{-1}$, from which \citet{2008A&A...487..223C}
inferred a distance of 1.6$\pm$0.4kpc. However, \citet{2010ApJ...724..480H} recently derived a distance of $630 \pm 100$~pc from space-based visual light curves that span the first 20 days after discovery. This much smaller distance estimate demands
that the VLTI-derived distance be scrutinized. Using the spatial expansion rate from
the VLTI but a revised velocity range of the dust-forming ejecta of 350$\pm$160\kms\ based on the mean
velocity measured from blue-shifted absorption lines of OI and
SiII, \citet{2012arXiv1203.6725N} derived a revised distance of 1.1$\pm$0.5kpc.

The dust event and the large amount of circumstellar material prevented any X-ray
detection of \object{V1280\,Sco} until May 2009, t=834d \citep{2009ATel.2063....1N}. After this date the 
source was detected several times from t=834d to t=939d after outburst \citep{2011ApJS..197...31S},
although the dense ejecta considerably absorbed the potentially important X-ray flux.

We report on observations of the Very Large Telescope performed between 2009 (t=857d)
and 2011 (t=1664d) that reveal a striking
bipolar nebula in expansion. The very slow ejecta of \object{V1280~Sco} likely expanded while being 
significantly influenced by the companion, although an intrinsically polar-oriented ejection
scenario must also be investigated. This is a unique opportunity to observe a newly formed bipolar nebula. Our study has a broader
relevance in the
context of the shaping of asymmetrical bipolar nebulae and the formation of disks, such as 
that seen in some planetary or pre-planetary nebulae \citep{2011apn5.confE.218C, 2011arXiv1102.4647D, 2011A&A...527A.105L, 2007A&A...473L..29C}, born-again sources 
\citep{2011MNRAS.410.1870L, 2009A&A...493L..17C} or in massive stars
\citep{2011A&A...526A.107M,2010A&A...512A..73M, 2010AJ....139.1993K, 2010ASPC..435..395K} 
and for a better understanding of the still poorly known common-envelope phenomenon
\citep{2011MNRAS.411.2277D}. 

The observations are presented in Sect.\,2. In Sect.\,3 we derive some physical
parameters of the nebula, and we discuss our results in Sect.\,4.

\begin{figure}
 \centering
\includegraphics[width=8.cm]{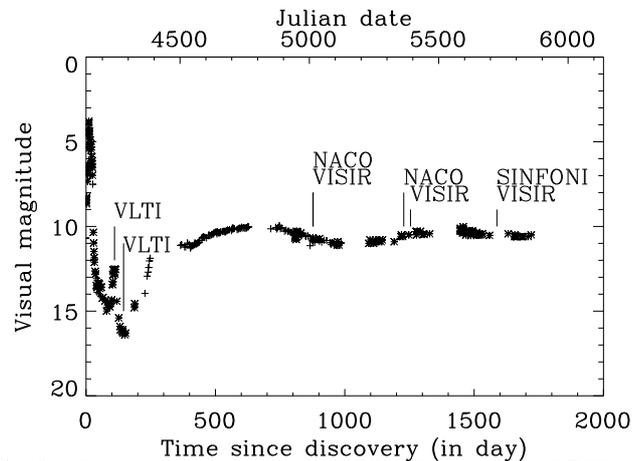}
 \caption[]{Visual light curve of \object{V1280~Sco} from AFOEV and ASAS-3
 data with the dates of the VLTI and VLT observations.  \label{fig:lightCurve}}
\end{figure}

\section{Observations}
\begin{table} 
  \begin{caption}
    {NACO/VLT, VISIR/VLT and SINFONI/VLT observing logs. 
     \label{tab:log_obs}
    } 
  \end{caption}
  \centering
  \begin{tabular}{lcc cc}
    \hline
    \hline
    Date &  day$^{\mathrm{1}}$ & SCI/CAL & Inst. & Filters \\
    \hline
    {\tiny 2009.07.02}  & 877 &V1280\,Sco & NACO & J, H, K\\
    {\tiny 2009.07.02}  & 877 &PSF 1$^{\mathrm{2}}$& NACO & J, H, K\\
    {\tiny 2009.07.03}  & 878 &V1280\,Sco & VISIR$^{4}$ & N band filters$^{5}$\\
    {\tiny 2009.07.03}  & 878 &PSF 2$^{\mathrm{3}}$ & VISIR$^{4}$ & N band filters$^{5}$\\
       \hline
    {\tiny 2010.06.17}  & 1228 &V1280\,Sco & NACO$^{4}$ & J, H, K \\
    {\tiny 2010.06.17} & 1228 &PSF 1$^{\mathrm{3}}$ & NACO$^{4}$ & J, H, K \\
    {\tiny 2010.07.28}  & 1254 &V1280\,Sco & VISIR$^{4}$ & N band filters$^{5}$\\
    {\tiny 2010.07.28}  & 1254 &PSF 2$^{\mathrm{3}}$& VISIR$^{4}$ & N band filters$^{5}$\\
       \hline
    {\tiny 2011.05.21}  & 1566 &V1280\,Sco & SINFONI &  H, K \\
    {\tiny 2011.06.13}  & 1589 &V1280\,Sco & VISIR$^{4}$ & N band filters$^{5}$  \\
    {\tiny 2011.06.13}  & 1589 &PSF 2$^{\mathrm{3}}$& VISIR$^{4}$ & N band filters$^{5}$\\
    {\tiny  2011.06.26}  & 1602 &V1280\,Sco & SINFONI &  K \\
    {\tiny 2011.06.27}  & 1603 &V1280\,Sco & SINFONI &  K \\
    {\tiny 2011.07.03}  & 1609 &V1280\,Sco & SINFONI &  H, K \\
    {\tiny 2011.07.03}  & 1609 &V1280\,Sco & SINFONI &  H, K \\
    {\tiny 2011.08.26}  & 1663 &V1280\,Sco & SINFONI &  H, K \\
    {\tiny 2011.08.28}  & 1664 &V1280\,Sco & SINFONI &  H, K \\

    \hline

 \end{tabular}
 	\begin{list}{}{}
 	\item[$^{\mathrm{1}}$] Day after outburst
	\item[$^{\mathrm{2}}$] GSC 07364-01316, J=8.29$\pm$0.03, H=7.68$\pm$0.07, 7.48$\pm$0.02 
	\item[$^{\mathrm{3}}$] HD152934, M5V, V=11, K=2.46, IRAS f12= 8.19 Jy
 	\item[$^{\mathrm{4}}$] Burst mode
	\item[$^{\mathrm{5}}$] PAH1, $\lambda$=8.59$\mu$m, $\Delta \lambda$=0.42$\mu$m; ArIII, $\lambda$=8.99$\mu$m, $\Delta \lambda$=0.14$\mu$m; 
PAH2, $\lambda$=11.25$\mu$m, $\Delta \lambda$=0.59$\mu$m; PAH2\_2, $\lambda$=11.88$\mu$m, $\Delta \lambda$=0.37$\mu$m;
NeII, $\lambda$=12.81$\mu$m, $\Delta \lambda$=0.21$\mu$m; NeII\_1, $\lambda$=12.27$\mu$m, $\Delta \lambda$=0.18$\mu$m
	\end{list}
 \end{table}

The source was observed with the NACO, VISIR and SINFONI instruments of the Very Large
Telescope (VLT). The log of the observation is shown in Table\,\ref{tab:log_obs} and the visible light curve is reproduced in Fig.\ref{fig:lightCurve}.

\begin{figure*}
 \centering
\includegraphics[width=6.9cm]{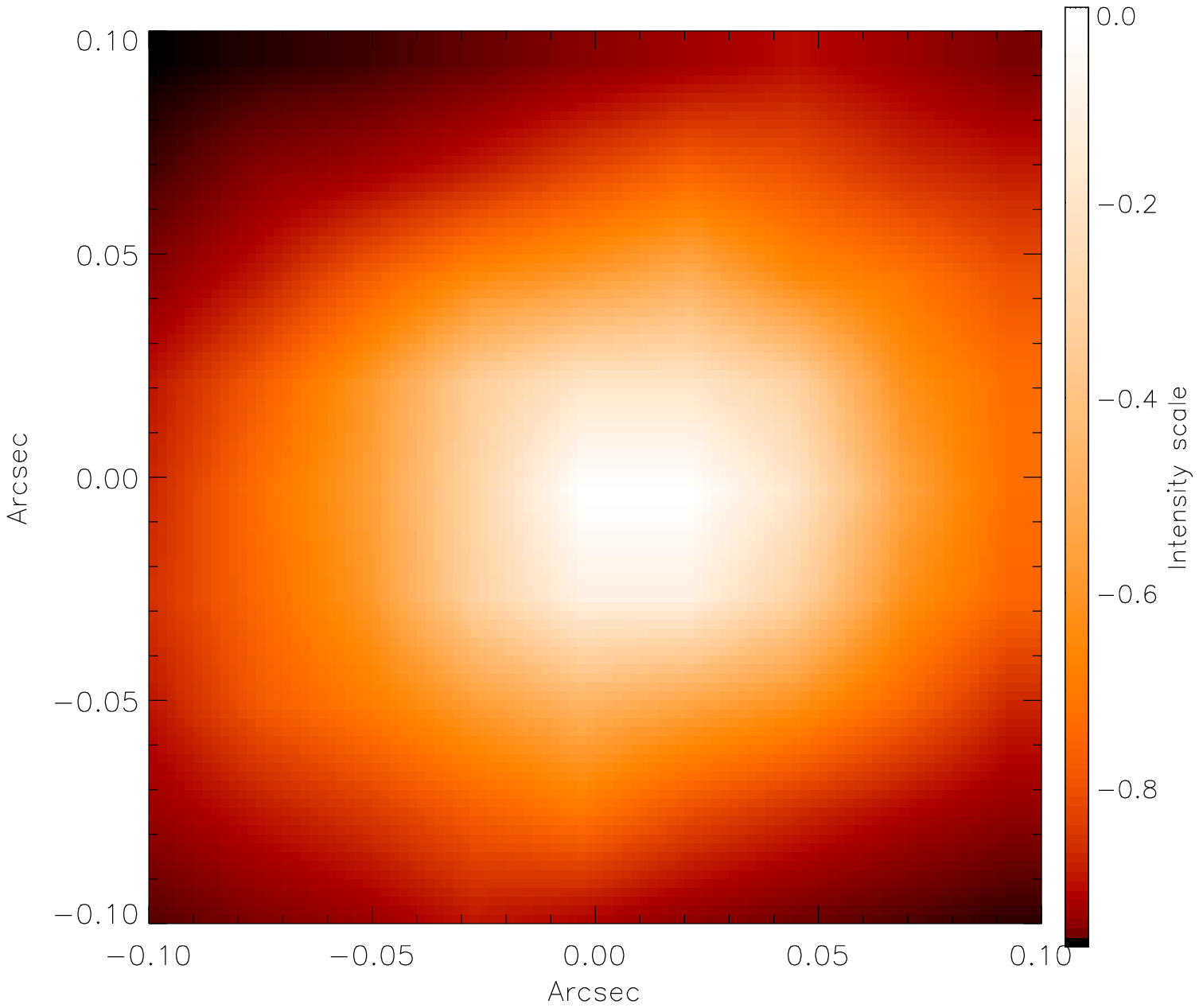}
\includegraphics[width=6.9cm]{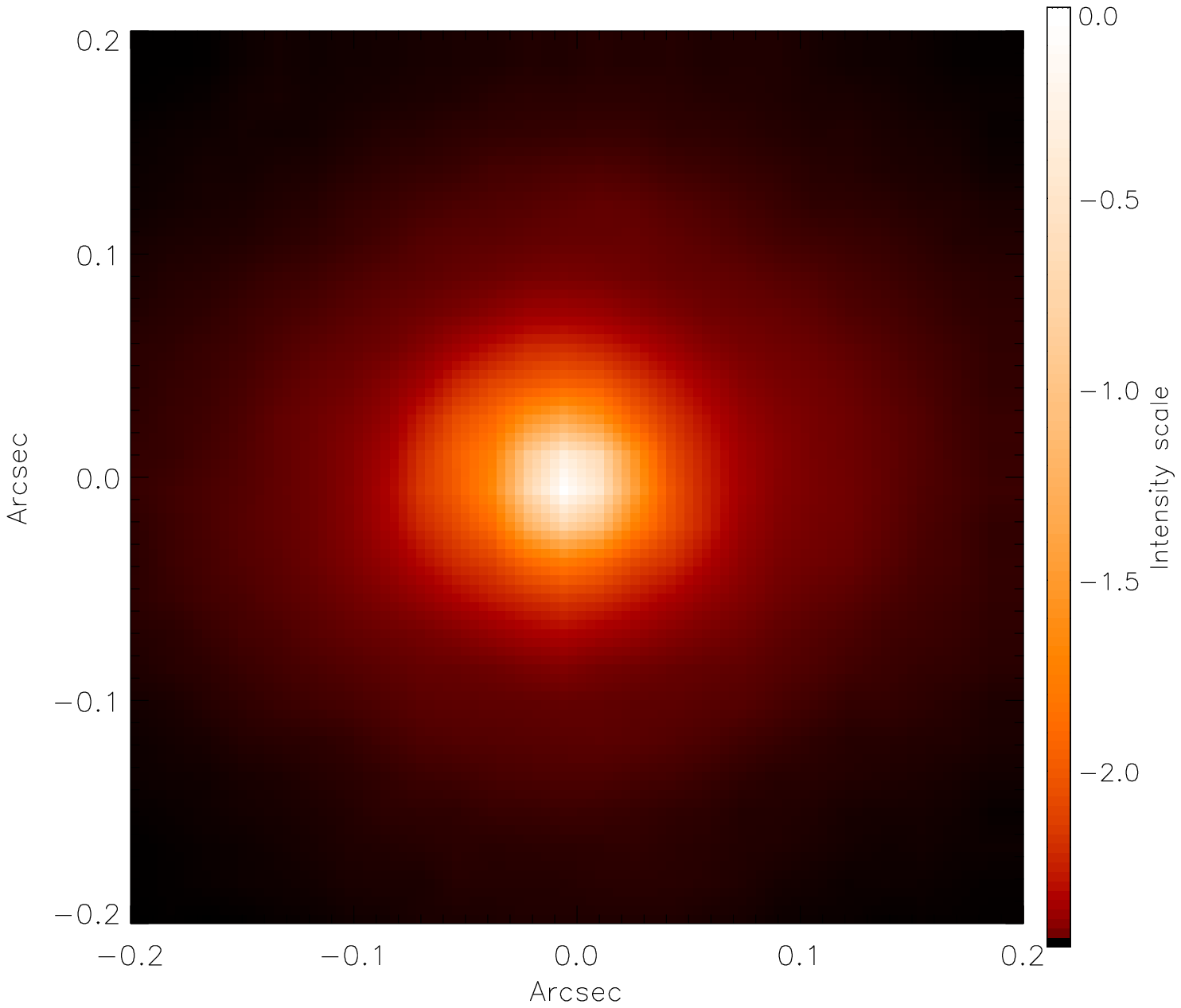}\\

 \caption[]{{\bf Left:} Normalized 2009 K band NACO image of V1280\,Sco in logarithmic scale. {\bf Right:} Same for the calibrator.
\label{fig:naco2009}}
\end{figure*}

\subsection{NACO observations}
We observed \object{V1280~Sco} in J, H, Ks with the S13 camera, which provides a
 field of view (FOV) of 14$\arcsec$x14$\arcsec$ and
a pixel scale of 13.26$\pm$0.03~mas. The
AutoJitter mode was used in which the telescope
moves according to a random pattern in a 5$\arcsec$ box at each exposure.
The nearby star GSC 07364-01316 \citep{2008AJ....136..735L} was selected as the reference star for
the visible wavefront sensor of the adaptive optics (AO) for determining the point spread function (PSF) of the instrument and to serve as a rough flux calibrator. The PSF obtained in 2009 with NACO in standard 
mode was not good enough to allow a meaningful deconvolution of the compact source. In 2010, the NACO observations benefited from the newly implemented
'cube mode', which provides a large set of short exposures (30ms) over part of the detector.
8000 frames were recorded per filter, representing an exposure time of 480s each.
This new capability is essential to reach truly 
diffraction-limited imaging at the shortest wavelengths \citep{2009A&A...504..115K}. 
The cube mode provides a large cube of short exposures from which a sub-set with the highest Strehl ratio is selected (principle of 'lucky imaging').


\begin{figure*}
 \centering
\includegraphics[width=6.9cm]{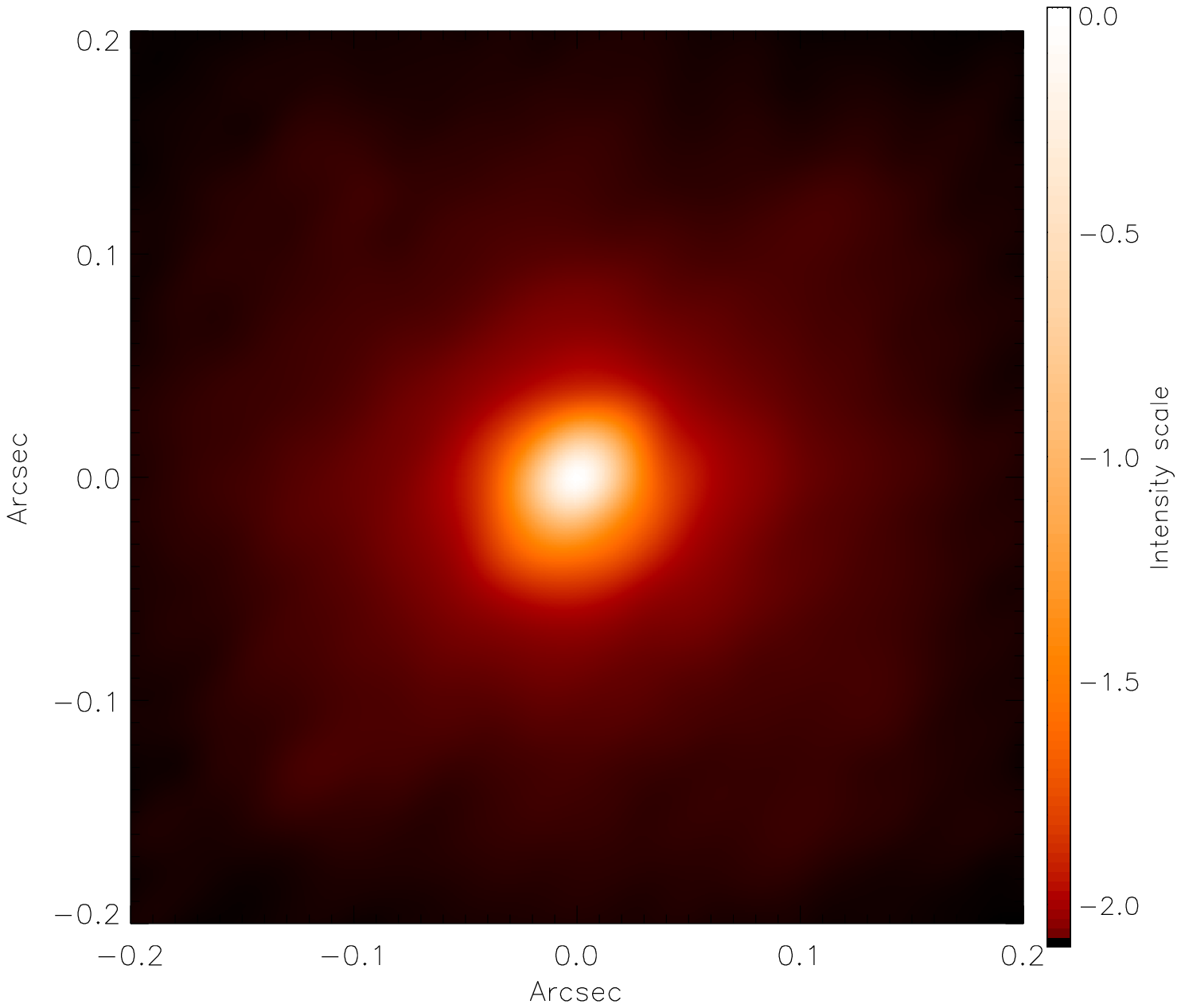}
\includegraphics[width=6.9cm]{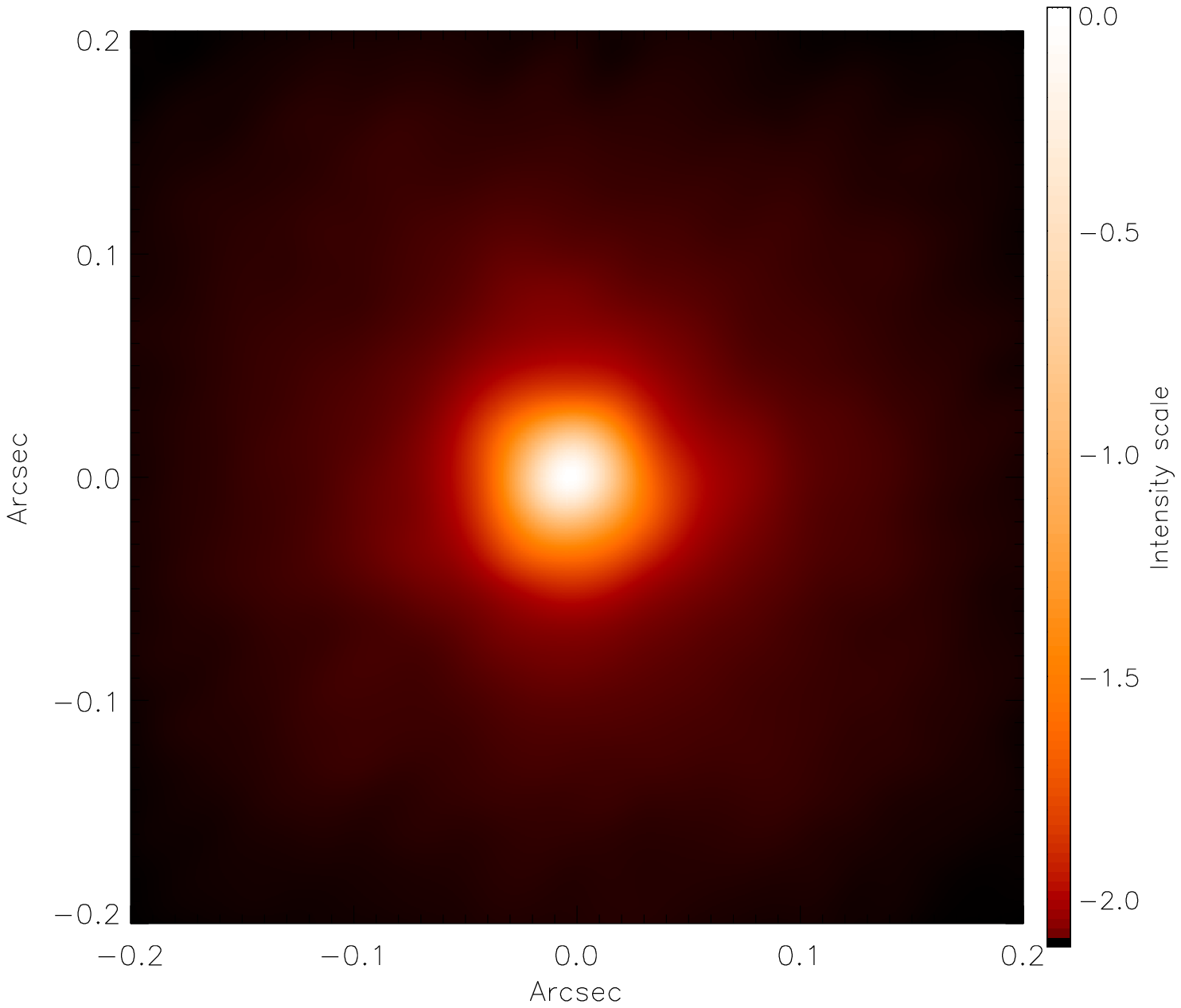}\\
\includegraphics[width=6.9cm]{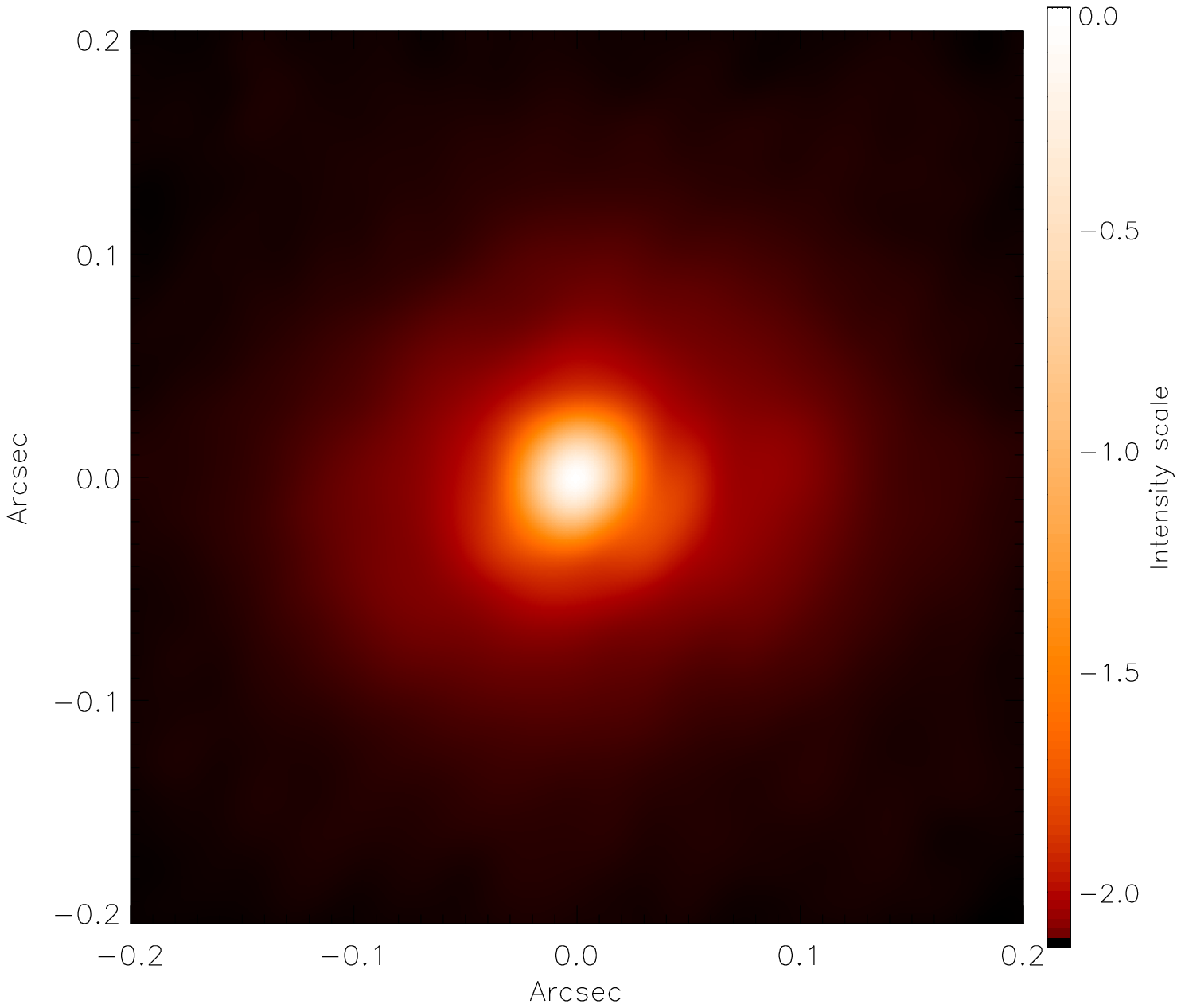}
\includegraphics[width=6.9cm]{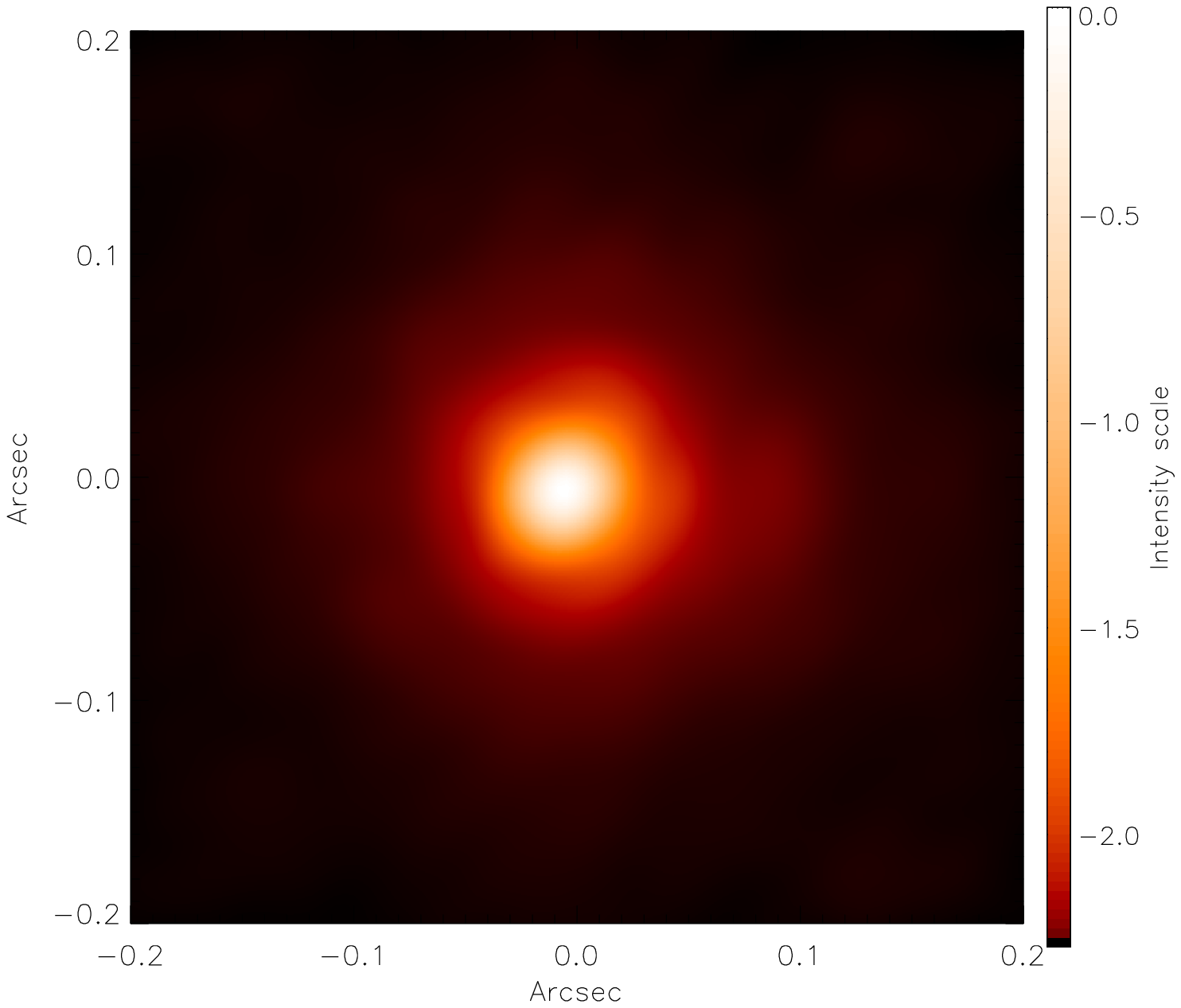}\\
\includegraphics[width=6.9cm]{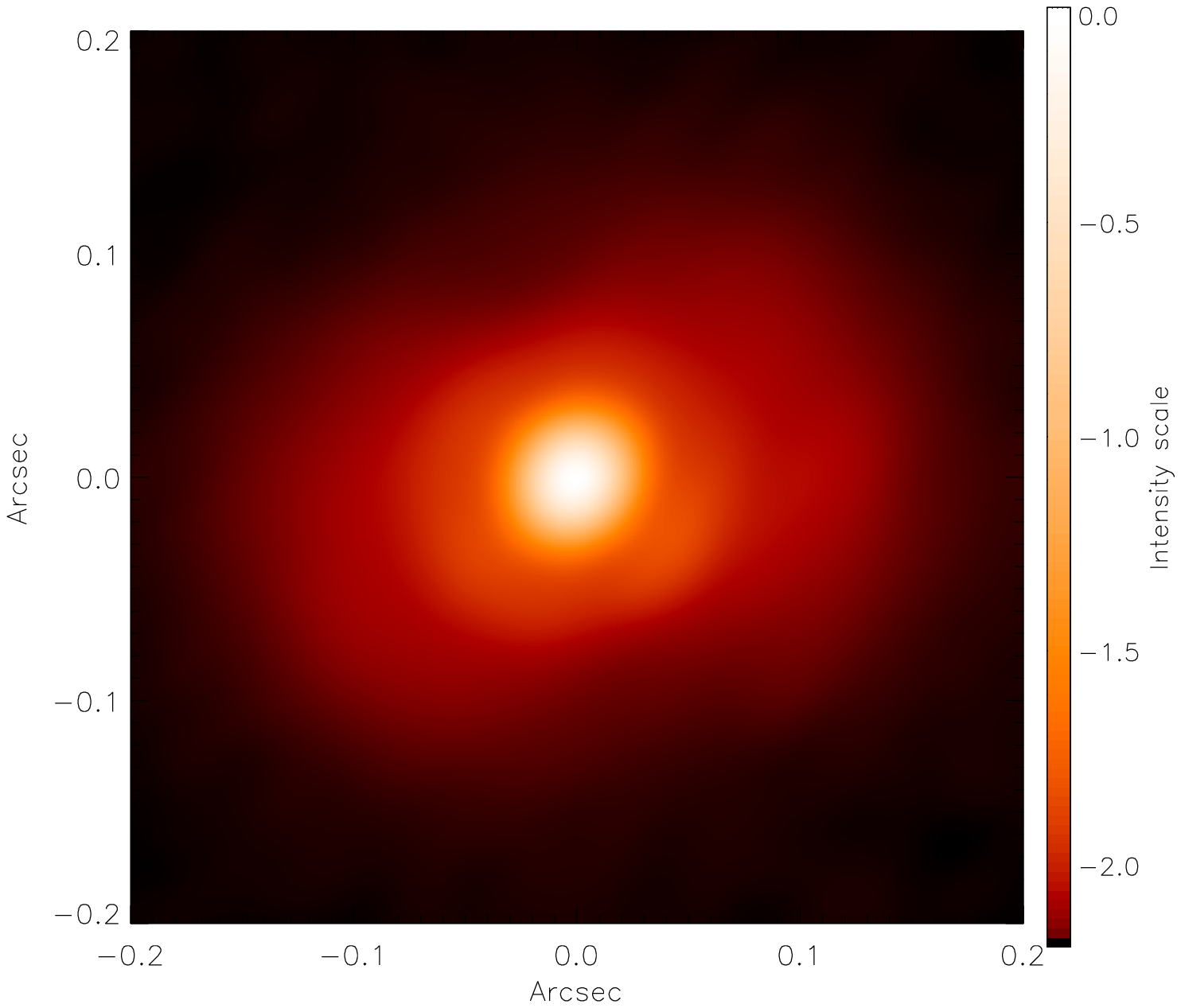}
\includegraphics[width=6.9cm]{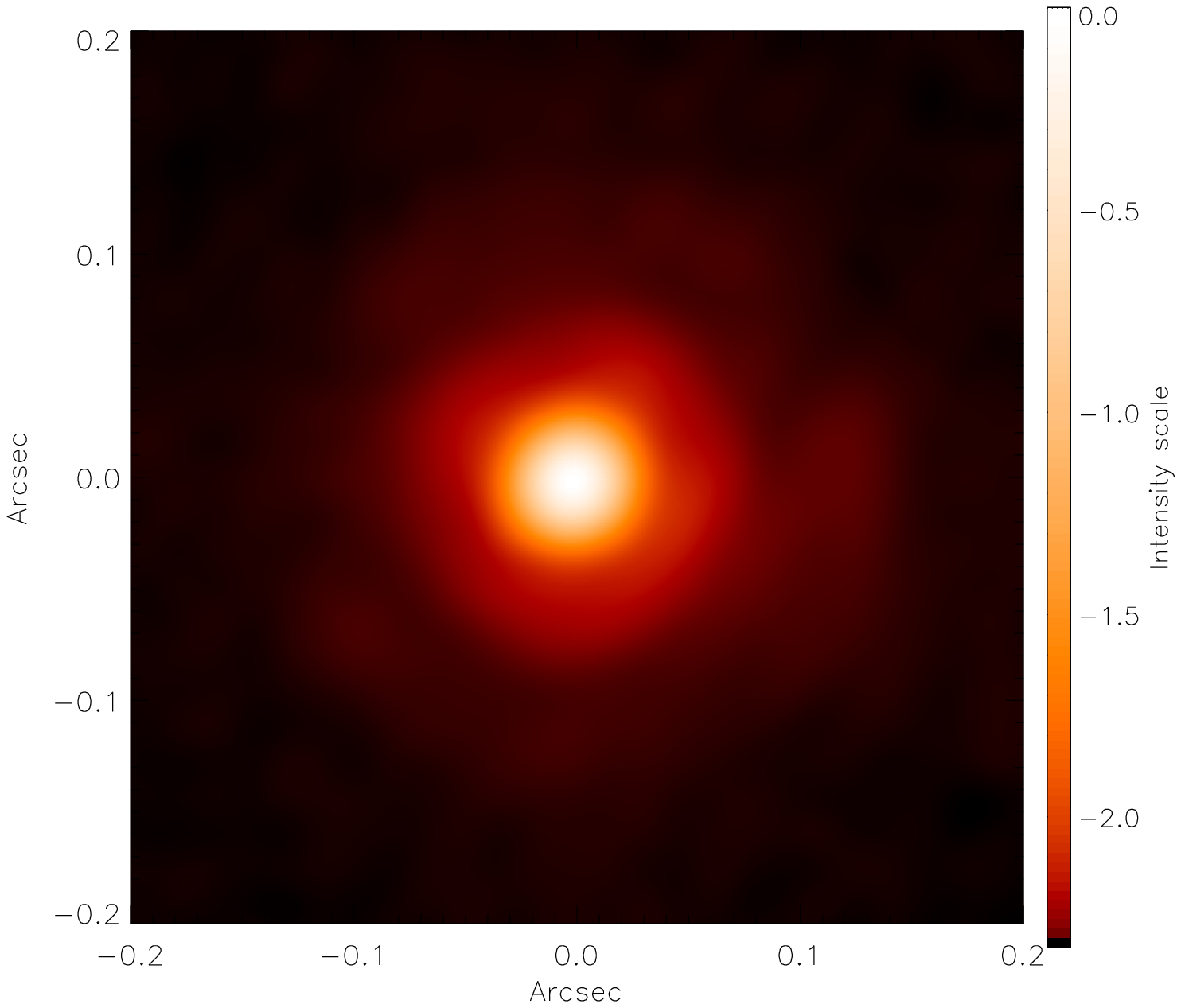}
 \caption[]{{\bf Upper left:} Normalized 2010 J band NACO image of V1280\,Sco in
 logarithmic scale. {\bf Upper right:} 2010 J band NACO image of the calibrator. 
 {\bf Middle:} Same in the H band. {\bf Bottom:} Same in the K band. A significant brightening of the central source (unresolved) is observed compared to 2009 (Fig.\ref{fig:naco2009}).

\label{fig:naco2010}}
\end{figure*}

\subsection{VISIR observations} The NACO observations were complemented by mid-infrared 
images obtained with the VISIR instrument \citep{2004Msngr.117...12L}. The VISIR observations were
performed in July 2009 (MJD\,55014.2 and 55015.1, t=877-878d), 
June and July 2010 (t=1228d and t=1254d) and June 2011 (t=1589d). The atmospheric conditions 
were stable for all nights. 
The imager was used in burst mode (DIT ranging from 10 to 25\,ms),
providing a pixel scale of 0.075\arcsec\ and a field of view of 19.2\arcsec\ x 19.2\arcsec. The chopping
throw was 8\,arcsec and the chopping frequency 0.25\,Hz. A set of narrow-band filters covering the full N band were used, with integrated total exposure times per filter ranging from 280s to 360s. Several
filters were centered on potential spectral features: PAH1, PAH2 and NeII. The exposures taken
with the PAH2 ($\lambda$=11.25$\mu$m, $\Delta \lambda$=0.59$\mu$m) and NeII filters ($\lambda$=12.81$\mu$m, $\Delta \lambda$=0.21$\mu$m) 
were immediately followed by exposures taken with a nearby continuum filter called
PAH2\_2 ($\lambda$=11.88$\mu$m, $\Delta \lambda$=0.37$\mu$m) and NeII\_1 ($\lambda$=12.27$\mu$m, $\Delta \lambda$=0.18$\mu$m).

The standard chopping/nodding technique
was applied to remove the sky contribution. The burst mode provides a large cube of short-exposure
frames in which all single chopping and nodding images recorded. 
The reduction of the VISIR data additionally includes the correction of various defects 
exhibited by the VISIR detector. 

 The VISIR burst mode and NACO cube modes are very similar. The principle is to acquire very short exposures
($\le$50\,ms) to keep the complete integration within a fraction
of the coherence time ($\sim$300\,ms at Paranal in the mid-IR). The burst modes of NACO and 
VISIR provided large cubes of short exposures that enable the reconstruction of 
quality-enhanced images using shift-and-add techniques, and selecting the best percentile (lucky imaging) following an 
algorithm described by \citet{2009A&A...504..115K} and \citet{2011MNRAS.417...32L}. 
The VISIR burst mode images of \object{V1280\,Sco} were deconvolved using a
Richardson-Lucy algorithm. Thirty iterations were sufficient to obtain good quality converged images.  The VISIR images are shown in Fig.\ref{fig:VISIR1} (2009), Fig.\ref{fig:VISIR2} (2010) and Fig.\ref{fig:VISIR3} (2011).

Flux calibration was performed using standard
aperture photometry methods applied to the science and reference
stars. The observations were performed in fairly similar atmospheric conditions and the
raw fluxes observed per filter using the same calibrator are similar at better than 5\%. Unfortunately, the VISIR PSF calibrator, \object{HD\,152934}, was chosen for its brightness but was a poorly suited photometric calibrator. \object{HD\,152934} is an anonymous S-type AGB star whose spectral type is not well constrained. We used synthetic templates of S-type stars produced by the MARCS codes \citep{2008A&A...486..951G} to fit the SED of \object{HD\,152934}. We then estimated the flux in the VISIR filters using the best synthetic spectrum ($T_{eff}=3000$\,K and $logg=0$) and calibrated the science object photometrically. This procedure is indirect and leads to a potential calibration bias larger than 10\% in some filters for such a complex source.

\subsection{SINFONI observations}

VLT/SINFONI is a near-infrared (1.1 - 2.45 $\mu$m) integral field spectrograph (IFS)
fed by an AO module \citep{2003SPIE.4841.1548E}. 
Integral field spectroscopy allows one to gather spectra of the sky over a two-dimensional 
field of view. Final products are data cubes with two dimensions of x, y and 
a third dimension formed by the wavelengths. During our observations the spectrograph operated in the \textit{K} and \textit{H} bands, 
providing a spectral resolution of 4000 and 3000, respectively. The FOV was cut 
into 32 slices, and each slice was projected onto 64 detector pixels. The field of view 
of the data set is 0.8\arcsec$\times$0.8\arcsec\ and is composed of spaxels of 12.5$\times$25~mas in the \textit{K} and \textit{H} bands.
We obtained several VLT/SINFONI integral field spectroscopic data sets for \object{V1280\,Sco} 
from May to August 2011. The FOV was oriented such that the major axis of the nebula
roughly coincided with the vertical direction. During that period, the central stellar-like source increased
considerably in flux, while the bipolar nebula in which this source is embedded had
a spatial extent of $\leq$0.7\arcsec\, which made these observations challenging. The observations were performed in service mode
and were repeated several times, but the atmospheric conditions never met the criteria for
providing good quality data. We nevertheless worked on the best data of the
sample and present the most robust results. The data set of 26 June 2011 exhibited the better signal-to-noise ratio (S/N) 
although the bipolar nebula was slightly cut in the north. Because of the limited S/N 
of the observations and the difficult removal of the telluric lines, a differential
approach, using a comparison between a given emission line flux and its corresponding adjacent continuum was employed in the reduction and analysis.

\section{Observational results}

\subsection{The dusty nebula}
A bipolar nebula with a high aspect ratio is clearly seen in the NACO, SINFONI and VISIR
images we obtained,  suggesting that the source is seen at high inclination. For our NACO and SINFONI data, the nebula 
extension was estimated by determining the regions of images with the highest 
gradient before reaching the noise level (or written differently, by the 
regions of densest contour levels before reaching the noise level). The 
estimation was performed on the images with and without PSF subtraction. The extensions measured
are reported in Table\,\ref{tab:ext}.

The near-IR images were dominated by the central source, which brightened considerably 
over the three years of observations.  The apparent size 
in the K band was 304 x 191 mas in 2009 (t = 877d).  Corrected for the PSF extension, this
leads to an angular size of 290$\pm$38 x 167$\pm$31 mas. After a PSF subtraction, the position angle (PA) of the major
axis is as $105 \pm 5^{\circ}$ with an axial ratio as $\sim 1.7$. However
a good estimate of the minor axis extension was difficult to obtain due to the bright central source,
and the artifacts remaining after the PSF subtraction.

The NACO observations performed in cube mode in 2010 (t = 1228d) were of excellent quality, and
the J, H and K bands can be fully exploited. An elongation was
clearly detected in each image at PA=105$\pm$3\deg, 102$\pm$3\deg and 107$\pm$3\deg. For instance,
the FWHM of the J band image of \object{V1280\,Sco} is 61$\pm$3mas 
along the major axis, and 54$\pm$3 along the minor axis. The apparent size in the K band was
404 x 260\,mas, which, corrected for the FWHM of the PSF, corresponds 
to 400$\pm$30 x 255$\pm$24 mas.

The 2011 SINFONI images (t = 1602d) suggest that the source is composed of a bright
unresolved source and two apparently well-detached polar caps. There seems to be 
no indication for extended equatorial material. The extension of the nebula in the
direction of the major axis is estimated to be 637$\pm$13mas, and the larger extent
of the caps perpendicular to the major axis to be 412$\pm$13mas (Fig.\ref{fig:SINFONI} and Table\,\ref{tab:ext}).

\begin{figure}
 \centering
\includegraphics[width=6.9cm]{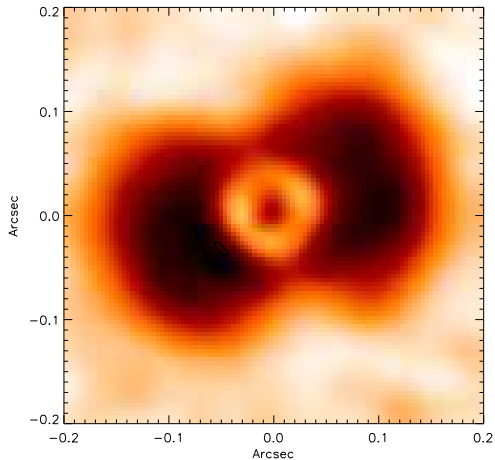}
 \caption[]{2010 NACO K band image after a PSF subtraction. The color table is inverted.
\label{fig:compContours}}
\end{figure}

The VISIR images also provided robust estimates of the PA at 108.5$\pm$2\deg\ in
2009, 109.1$\pm$1.6\deg\ in 2010 and  108.9$\pm$2.5\deg\ in 2011.
The bipolar morphology of our VISIR images was more clearly seen in the most recent
images due in part to the expansion of the nebula. Fig.\ref{fig:expVISIR} (left) shows a cut
of the VISIR images obtained at 8.59 $\mu$m in 2009, 2010 and 2011,
together with a PSF for comparison. This filter provides the shortest wavelength,
hence the best spatial resolution of VISIR and was therefore used for providing the extension estimates (Table\,\ref{tab:ext}). The bipolar lobes of \object{V1280\,Sco} are  
clearly expanding with time, and in 2011, one could see 
a double peaked cut generated by the two emission peaks of the bipolar nebula.
One-dimensional Gaussian fitting was performed to measure the FWHM of the
cuts, and the results are shown in Table\,\ref{tab:ext}. After deconvolution, the VISIR images provide a size and an axial ratio equivalent
to that seen in the near-IR. The size and appearance of the nebula extension was nearly unchanged from
8 to 13 micron. The deconvolved images confirm that the object is bipolar and
expanding. Our most resolved images (from our 2011 dataset) show that the intensities of the two bipolar lobes are different, the southern lobe is brighter. This is probably due to an absorption effect as a consequence of this lobe being closer to us than the northern one. No mid-IR emission is seen between the caps. Fig.\ref{fig:expVISIR} shows a radial cut along the poles at 8.59. 8.99, 11.25, and 12.81 $\mu$m.
The separation between the two lobes does not depend on the wavelength. The ratio of the intensity
between the two peaks increases with wavelength due to absorption and optical depth effects.

\begin{figure*}
 \centering
\includegraphics[width=12.cm]{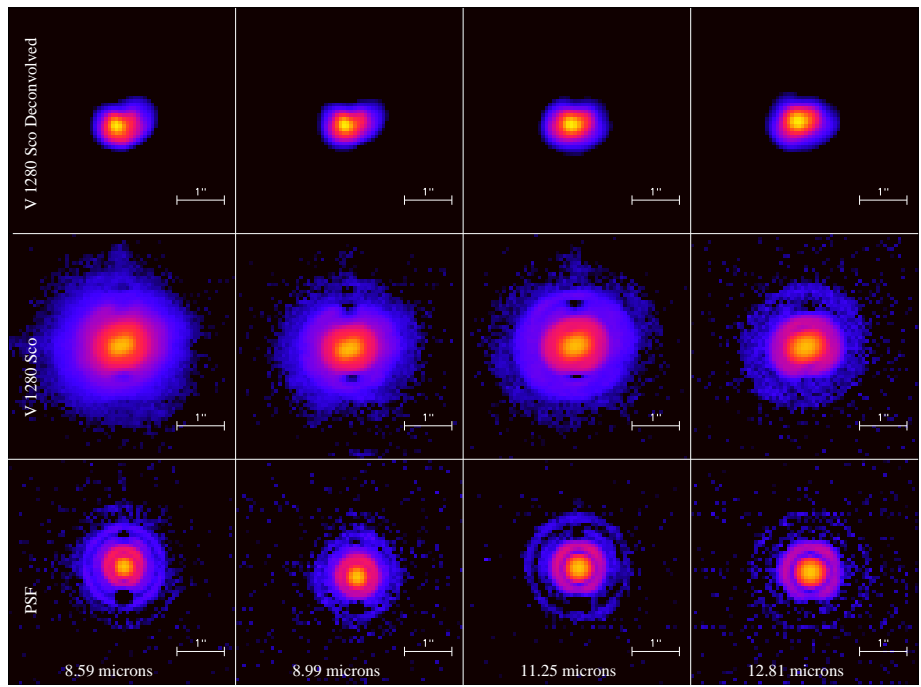}
 \caption[]{Burst-mode VISIR observations performed in 2009 (t=878d) with four different filters (see description in Table\,\ref{tab:log_obs}). {\bf Up:} Deconvolved images of \object{V1280\,Sco}. {\bf Middle:} Mean images generated from the best short exposures of \object{V1280\,Sco}. {\bf Bottom:} Mean images of the PSF with the same treatment.
\label{fig:VISIR1}}
\end{figure*}

\begin{figure*}
 \centering
\includegraphics[width=12.cm]{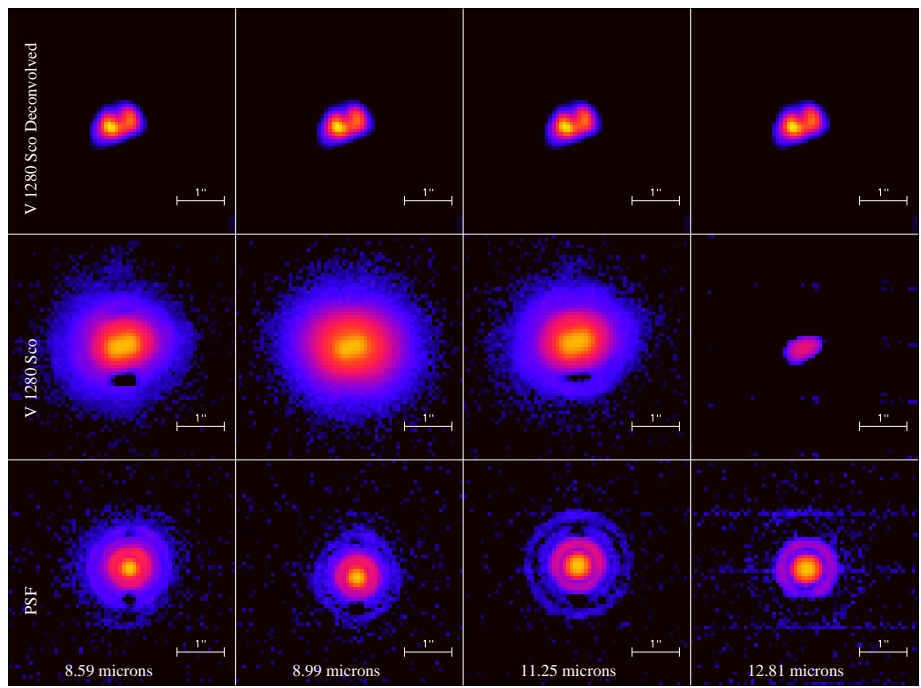}
 \caption[]{Same for the VISIR observations performed in 2010 (t = 1254d) 
\label{fig:VISIR2}}
\end{figure*}

\begin{figure*}
 \centering
\includegraphics[width=12.cm]{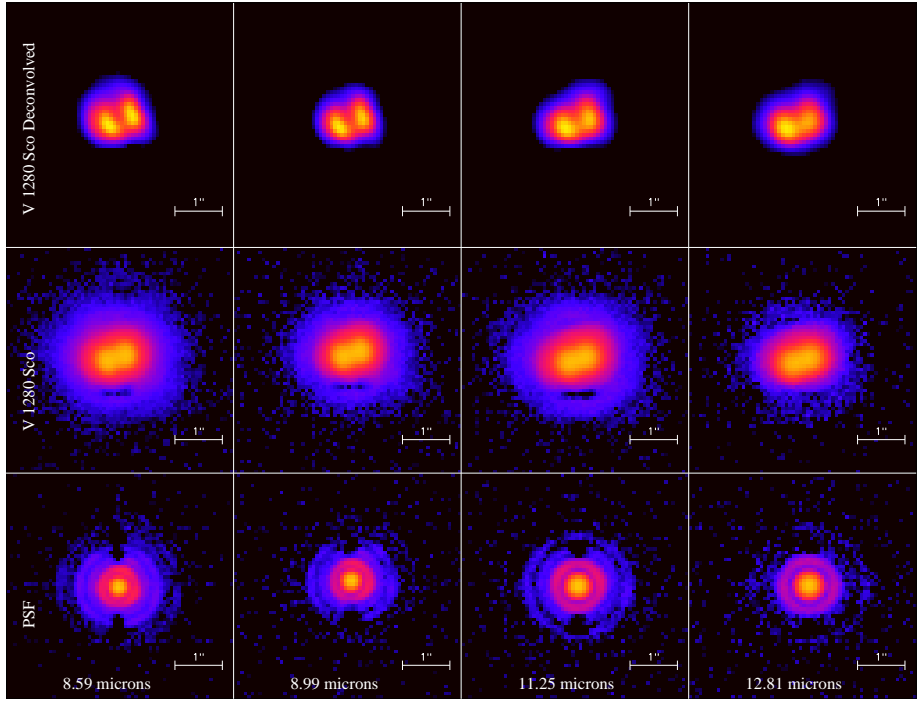}
 \caption[]{ Same for the VISIR observations performed in 2011 (t = 1589d)
\label{fig:VISIR3}}
\end{figure*}

\begin{table} 
  \begin{caption}
    {Extensions of the bipolar nebulae. 
     \label{tab:ext}
    } 
  \end{caption}
  \centering
  \begin{tabular}{lccc}
    \hline
    Filter & Major axis$^{\mathrm{1}}$& Minor axis$^{\mathrm{1}}$  & PSF FWHM\\
    \hline
    \multicolumn{4}{c}{2009}\\
    \hline
    K & 0.29$\pm$0.04\arcsec & 0.17$\pm$0.03\arcsec&0.092$\pm$0.005\arcsec\\
    PAH2 & 0.31$\pm$0.04\arcsec& $-$&0.23$\pm$0.03\arcsec\\
    \hline
    \multicolumn{4}{c}{2010}\\
    \hline
    K &0.40$\pm$0.03\arcsec & 0.26$\pm$0.02\arcsec  &0.057$\pm$0.003\arcsec\\
    PAH2 & 0.49$\pm$0.05\arcsec & 0.33$\pm$0.05\arcsec &0.23$\pm$0.03\arcsec\\
    \hline
    \multicolumn{4}{c}{2011}\\
    \hline
    K & 0.62$\pm$0.05\arcsec & 0.39$\pm$0.01\arcsec &0.13$\pm$0.02\arcsec\\
    PAH2 & 0.64$\pm$0.06\arcsec &0.37$\pm$0.05\arcsec &0.23$\pm$0.03\arcsec\\
    \hline
 \end{tabular}
 	\begin{list}{}{}
 	\item[$^{\mathrm{1}}$] Corrected for the FWHM of the PSF 
	\end{list}

 \end{table}


\subsection{Spectroscopy and flux measurements}
\label{sec:spectro}

The V flux has reached a roughly stable plateau 
at a magnitude of about 10.5 over a long period \citep{2012arXiv1203.6725N} and the central source radiates  efficiently in the K band. As shown by \citet{2009AAS...21442808L}, the nova is still powering a strong wind, and extensive nuclear burning is thus thought to have continued. 

An unresolved point source dominates the near-IR observations, 
whereas no point source contribution is observable in the mid-IR for the contemporaneous VISIR measurements. In 2009, the
stellar source (i.e. the unresolved flux) accounts for 54$\pm$6\%, increasing in 2010 
to reach 76$\pm$6\% (NACO data) and in 2011 87$\pm$5\% (SINFONI) of the total flux. These numbers are potentially 
biased by the different Strehl ratio reached by the AOs during the observations. 
These estimates were also possible in the H and J bands for the 2010 NACO observations, with 96$\pm$3\% and
100$\pm$3\%, respectively. This increase of the central source contribution indicates a
decrease of the circumstellar absorption, related probably to the expansion and dissipation 
of the nearby dust.

\begin{figure*}
 \centering
\includegraphics[width=8.5cm]{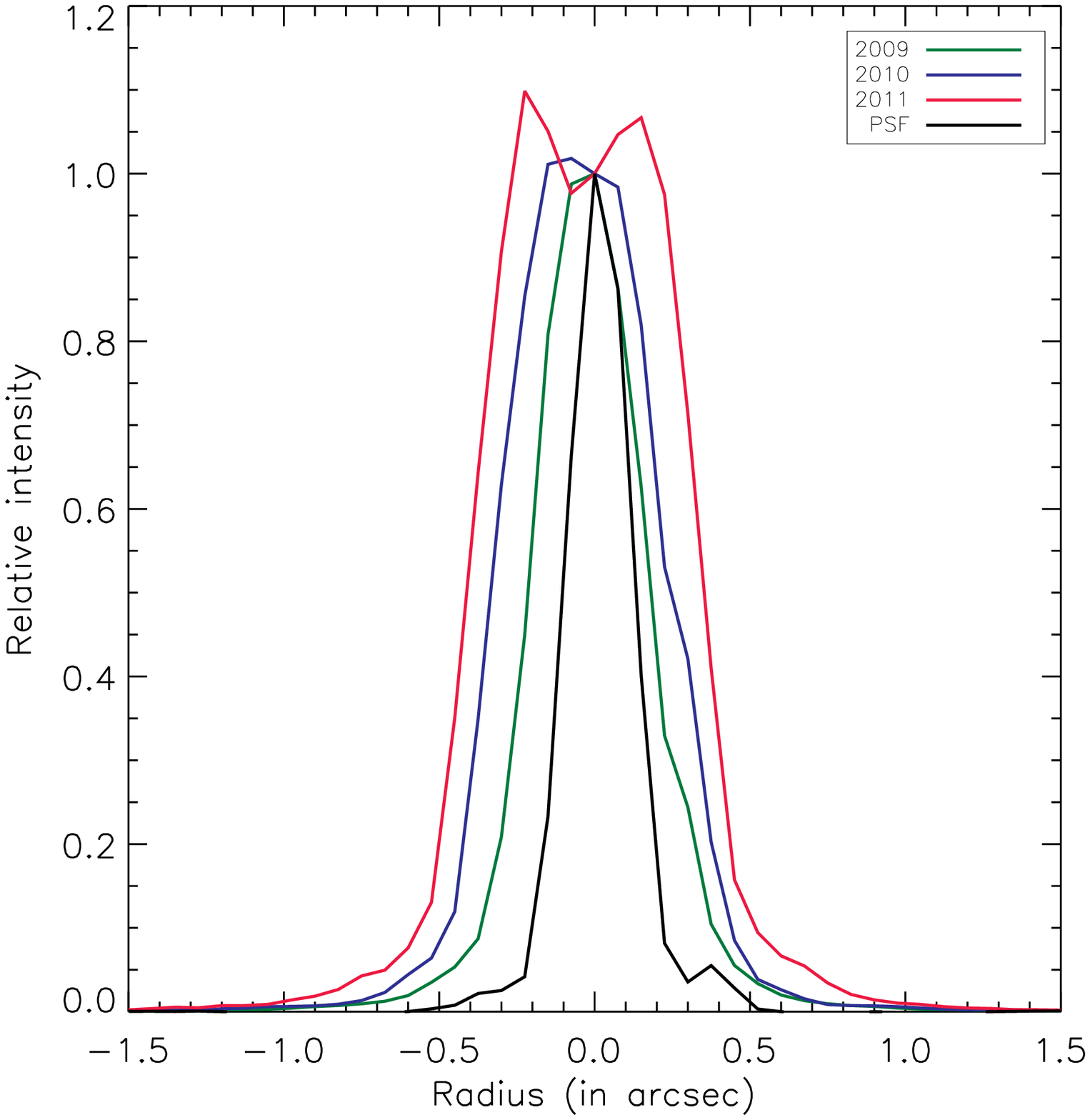}
\includegraphics[width=8.5cm]{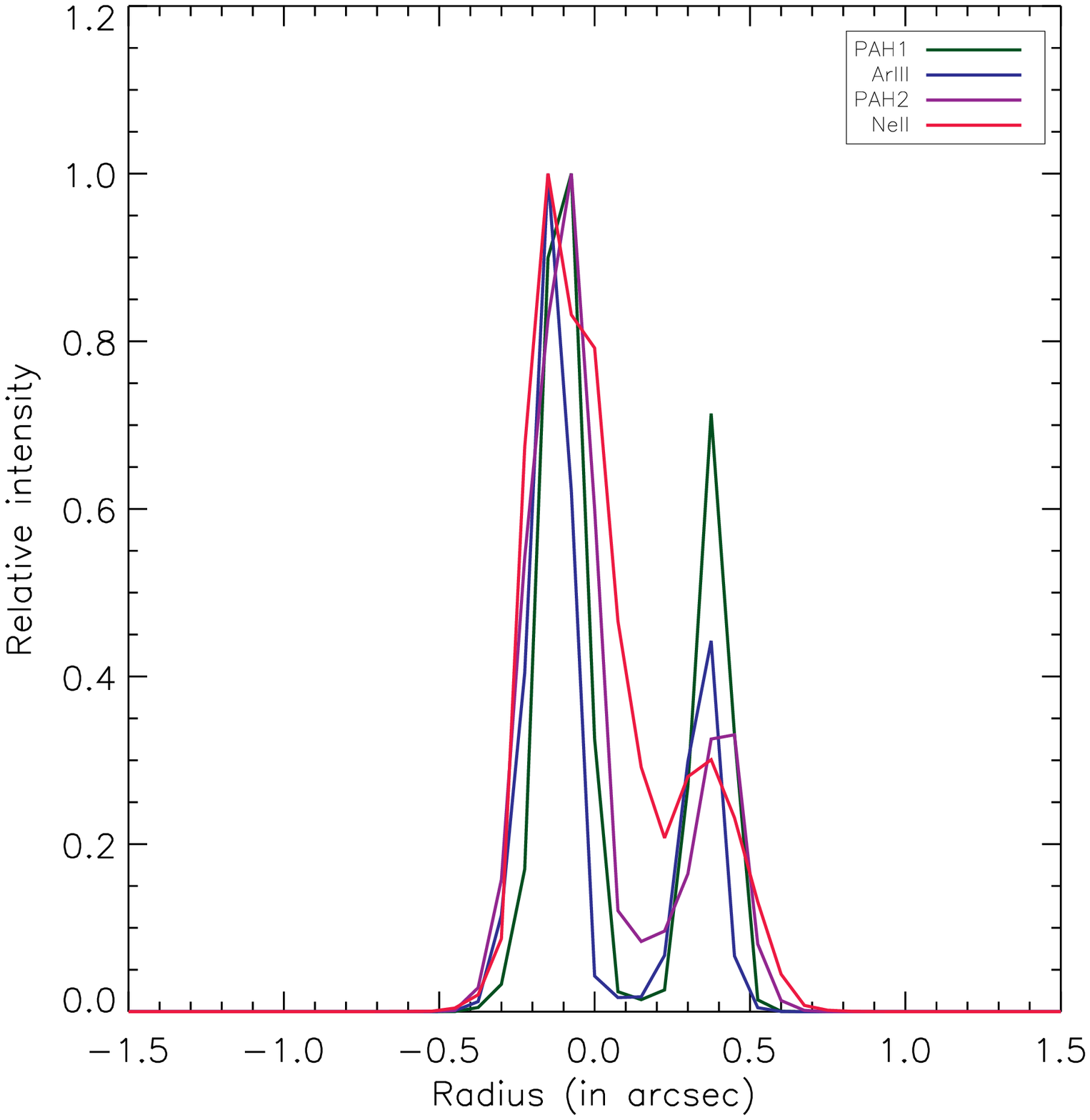}
 \caption[]{\label{fig:expVISIR} Left: Cut of the VISIR raw images at 8.59 $\mu$m along the major axis of the dust shell
 for the three epochs.
 An expansion of the dusty material is clearly seen. Right: Cut of the VISIR 2011 deconvolved images 
 along the main axis of the dust shell for all filters.} 

\end{figure*}


The photometry of the central source and its nearby environment was performed from the NACO, 
SINFONI and VISIR images and is provided in Table\,\ref{tab:phot}. These instruments or the mode used are not 
necessarily optimized to provide accurate photometry but despite the level of errors such 
information remains of high interest for these variable sources.

Many lines are detected by SINFONI (t=1609d). Of importance is the simultaneous detection of relatively 
high-excitation lines (mostly from the Balmer series), and the detection of low ionization 
species such as NaI and MgI, as already reported by \citet{2008MNRAS.391.1874D} 
and \citet{2010PASJ...62L...5S} for this source. The mean FWHM of the Balmer lines is 162$\pm$14\kms, and we note that Br$\gamma$ exhibits a broad pedestal ($\sim$700\kms). 
Some lines of low excitations are observed, the NaI\,2.149 line (FWHM=110$\pm$18\kms), 
MgI\,1.711 (FWHM=144$\pm$8\kms), some are unidentified such at 2.137$\mu$m (FWHM=105$\pm$8\kms) 
and some with weak signatures such as MgI\,1.576, CI\,1.690, as reported by \citet{2008MNRAS.391.1874D}. These lines are narrow and almost unresolved given the spectral resolution of these observations (R$\sim$3500 corresponding to a FWHM of 85\kms). The presence of the NaI\,2.149 line implies that 
the NaI must be protected from the central source, and is therefore 
very likely located within the dust. The presence of these low-excitation lines so late (t=1609d) after the outburst highlights how slowly \object{V1280\,Sco} has evolved, because these lines are not expected at this stage.

\begin{figure*}
\includegraphics[width=5.6cm]{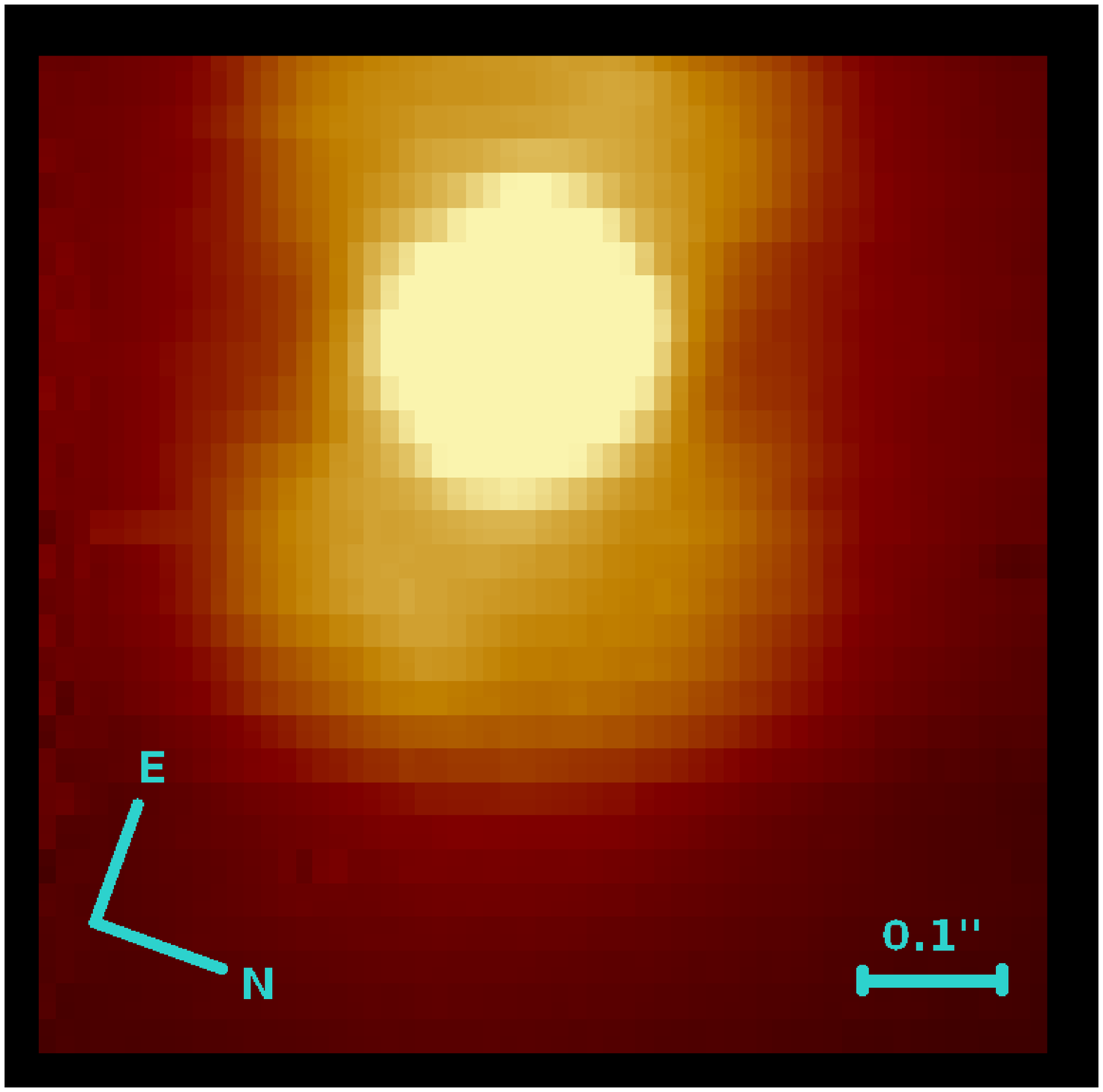} 
\includegraphics[width=5.6cm]{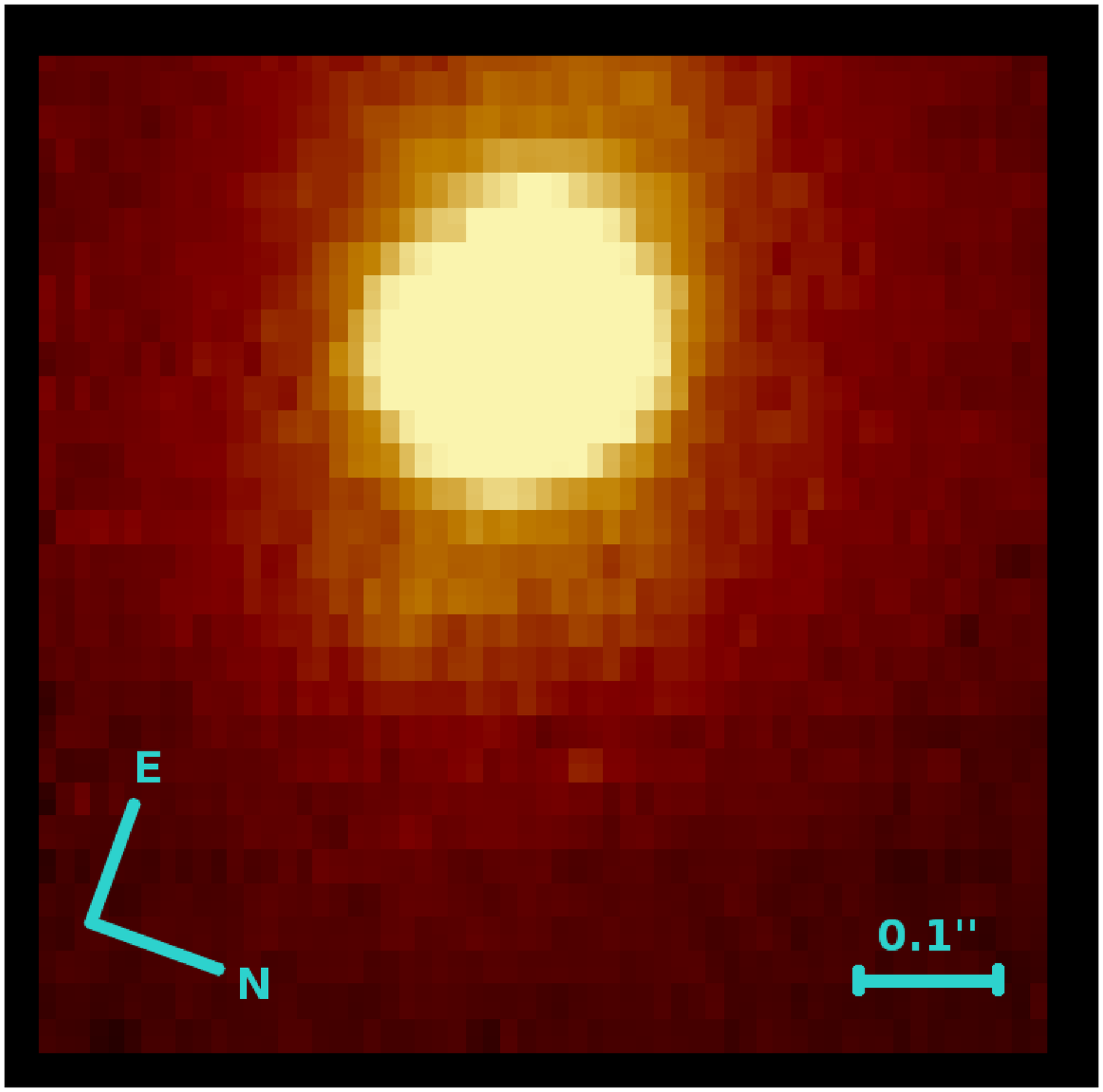} 
\includegraphics[width=5.6cm]{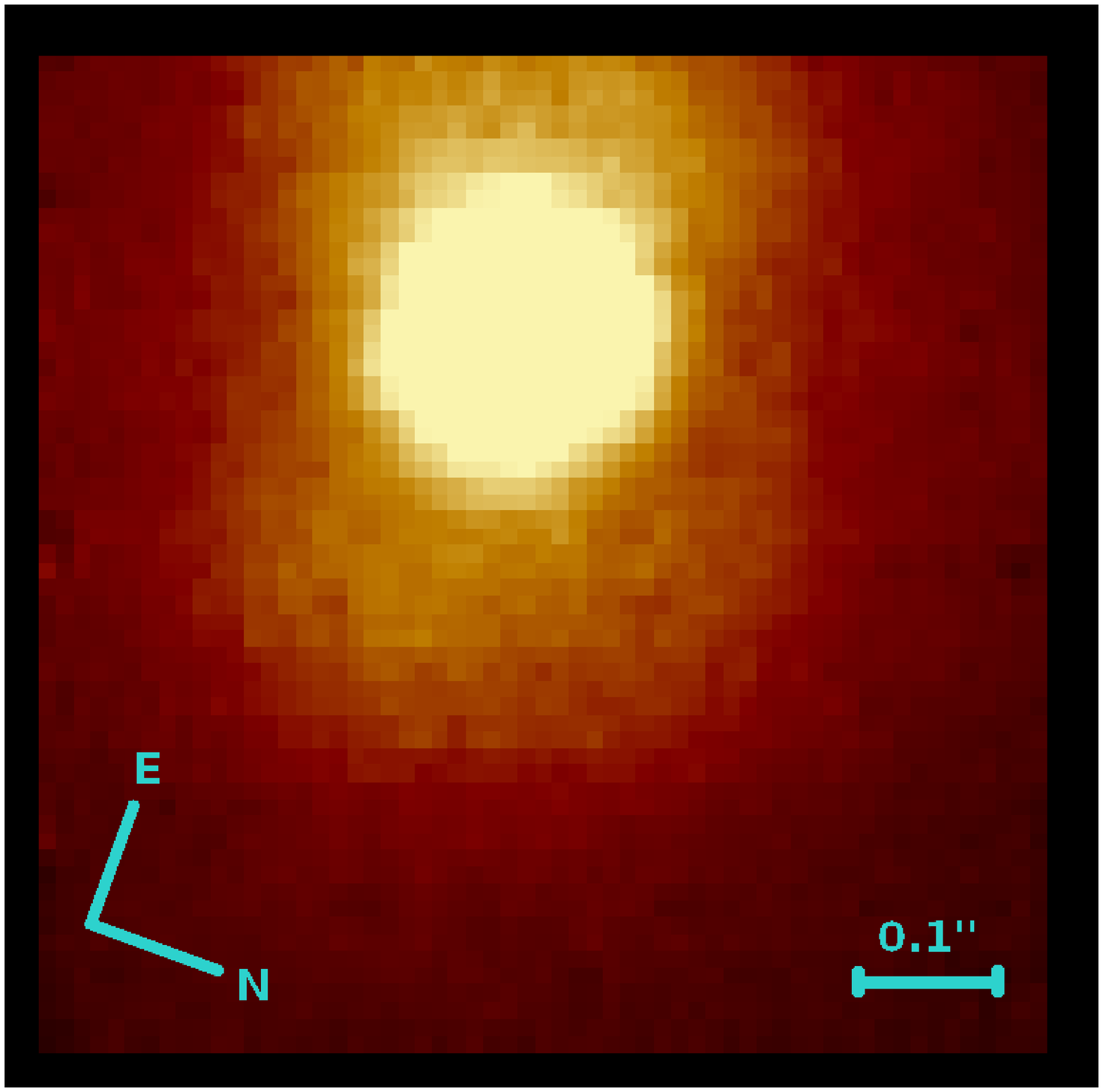}
\caption{{\bf Left}: Continuum flux from a narrow spectral channel (2.7nm) close to the Br$\gamma$ line. {\bf Middle}: same from a channel centered on the Br$\gamma$ line emission from which the nearby continuum shown  on the left was removed. {\bf Right}: same from a channel centered on the HeI2.05 line emission isolated from a nearby continuum. The Br$\gamma$ line forming region is centered on the outbursting source, whereas some intrinsic HeI2.05 emission seems to originate from the lobes. \label{fig:SINFONI} }
\end{figure*}

For the strongest lines, a SINFONI image was created by integrating the flux within a 
narrow spectral range, and by subtracting a red- and blue-sided continuum. For the
Br$\gamma$ line, the main source of the emission is observed in the central object, 
which contributes 98$\pm$7\% of the total flux. A slightly different behavior may be observed for 
the He~I~2.06\,$\mu$m (see Fig.\ref{fig:SINFONI}) line for which the core contributes 93$\pm$7\%,
and for which the lobes seem better contrasted. In the case of \object{V1280\,Sco},
one may hypothesize that the He~I~2.06\,$\mu$m emission in the caps could arise either from a 
shock between the fast nova wind and the slowly expanding, dust rich polar caps, 
or by a radiative interaction between this material and the growing UV flux from the white dwarf.   

The PAH\_2 filter is designed to detect carbonaceous molecules. This filter is not best suited because the 11.25$\mu$m feature in novae is seen at 
$\sim$11.4$\mu$m \citep{2005MNRAS.360.1483E}, although most of the flux should be isolated by the filter given the bandwidth (0.59$\mu$m). Yet, it is challenging to detect any morphological difference between images in the two filters because of the limited spatial resolution and the weakness of the flux 
contrast. It is not evident that the regions of PAH emission is distinct from that of the continuum. Spitzer IRS (6 May 2008) and IRTF/Spex 
(20 May 2009) observations show that dust emission features appeared at 3.28, 3.4, 
8.3, 8.7, and 11.4 microns, the so-called UIRs (unidentified infrared bands). Likely, carbonaceous molecules also formed in situ from condensed dust \citep[HAC]{2009ApJ...704..226B, 2010MNRAS.406L..85E, 2005MNRAS.360.1483E}.

\begin{table} 
  \begin{caption}
    {Photometry of the total source. 
     \label{tab:phot}
    } 
  \end{caption}
  \centering
  \begin{tabular}{lccc ccc}
    \hline
    Filter & $\lambda_0$ & t = 878d  & t = 1254d & t = 1589  \\
     & [micron] & [Jy] & [Jy] & [Jy]\\
     \hline
    J & 1.05 & - & 0.38$\pm$0.07& - \\
    H & 1.65 & - & 0.44$\pm$0.04 & -\\
    K & 2.1 & 0.67$\pm$0.04 & 0.44$\pm$0.02 & -\\
    PAH 1 & 8.59 & 47$\pm$6& 34 $\pm$ 4 & 28$\pm$ 4\\
    ArIII & 8.99 & 43$\pm$5& 32 $\pm$ 4 & 26 $\pm$ 3\\
    PAH 2 & 11.25 & 33$\pm$3& 25$\pm$3 & 20$\pm$3 \\
    PAH 2\_2$^{\mathrm{1}}$ & 11.88 & 31$\pm$4& 23$\pm$3 & 19$\pm$3 \\
    NeII & 12.81 & 26$\pm$3& 20$\pm$3 & 16$\pm$2\\
    \hline
 \end{tabular}
 	\begin{list}{}{}
 	\item[$^{\mathrm{1}}$] Filter dedicated to continuum measurements close to the PAH2 filter.
	\end{list}

 \end{table}

%




\subsection{The expansion rate}
In the near-IR, the datasets are not homogeneous with images taken with NACO in 
2009 without cube mode, then in 2010 in cube mode and in 2011 using the SINFONI
instrument. We did not attempt to derive an expansion rate for the lobes in the K band due to the variances in the observational techniques. Despite their lower spatial resolution, the images from the 
VISIR instrument were obtained solely in burst mode. Moreover, in the mid-IR, the central source is too faint to be detected, and the 
images probe the contribution from the dust alone. The NACO and SINFONI images exhibit a bright
stellar contribution that has to be carefully taken into account and removed before any geometrical information can be inferred.
The expansion was measured from the raw and deconvolved VISIR images. 
Fig.\ref{fig:expVISIR} shows the cuts of the images along the direction 
of the major axis and the results are shown in Table\,\ref{tab:ext}. Correcting the FWHMs from the
FWHM of the PSF at 8.59$\mu$m, this leads to an estimate of the expansion rate of 0.36$\pm$0.06 (2009),
0.39$\pm$0.05 (2010) and 0.40$\pm$0.04 (2011)\,mas per day, from which we infer a global estimate
of 0.39$\pm$0.03 mas per day. As an independent test, one can compute the separation 
of the polar caps as measured by SINFONI to be 0.62$\pm$0.05\arcsec at day 1602, providing an expansion rate  very close to the VISIR measurements.

This value, measured in the direction of the major axis of the nebula is close
to the expansion rate estimate of 0.35$\pm$0.03\,mas per day from the VLTI in a
similar wavelength range. The VLTI baselines used to infer
the expansion of the early dusty envelope were in a direction close to the major
axis of the nebula (within typically 30\deg), and therefore approximately probed 
the velocities projected on the sky in that direction. The bias expected from the use of baselines
not aligned to the major axis is an underestimate of the expansion rate.

\section{Discussion}

\subsection{The distance and expansion velocities}
Using the angular expansions measured from the infrared images, one can try to improve upon the distance estimate for \object{V1280\,Sco}.
The angular expansion inferred from the VLTI observations was 0.35$\pm$0.03 mas/d,
while the high-quality VISIR images obtained in the study suggest a slightly
higher expansion rate {\it in the direction of the major axis of the nebula} of 0.39$\pm$0.03 mas/d.

Combined with a measure of the ejecta expansion velocity, the angular expansion rate (mas/d) can be used to derive a distance to \object{V1280\,Sco}, under the hypothesis of a spherical ejection following

\begin{equation}
D_{\rm{kpc}} = 1.154 \times 10^{-3} \frac{V_{ej}}{\frac{d\theta}{dt}},
\label{formula1}
\end{equation}

\noindent where $V_{ej}$ is
the velocity of the ejecta in  \kms\ and $\frac{d\theta}{dt}$ is the rate of expansion in
mas per day.

Assuming a velocity $V_{ej}$ for the ejecta of 500$\pm$100km.s$^{-1}$, \citet{2008A&A...487..223C} estimated the distance to be 1.6$\pm$0.4 kpc. 

The difficulty of this exercise is to ensure that the Doppler velocities inferred
from the emission lines can be linked to the dust expansion measured in the VLTI data or VISIR images. These quantities must be measured at
the same wavelengths, if possible centered on an emission line that provides the kinematics. Moreover,
the Doppler velocities are translated into ejection velocities assuming a projection factor 
(the "p" factor) whose value depends on the kinematics but also on the flux distribution
of the source. 

The discovery of the strong bipolarity of the source implies a complex velocity field
that cannot be directly related to a limited number of velocity measurements. The 
expansion velocities were measured in the direction of the major axis of the bipolar
nebula and we adopted a Doppler velocity of 500km.s$^{-1}$, based 
on several radial velocity measurements from various emission lines. Co-mixing of the dust and gas emitting regions cannot be unequivocally demonstrated, and it is unknown whether the line-forming region coincides with the dust-continuum-forming region. Therefore, it is 
very difficult to measure the ejection velocity of the dusty nebula.

\citet{2010ApJ...724..480H} derived a much closer distance estimate, 630$\pm$100~pc
based on various estimates of intrinsic luminosity of the source. However, despite the quality of their photometry, many parameters are used to
infer the distance, some of them raised to high power indices, implying potentially 
large errors. \citet{2010ApJ...724..480H} found Mv = -6.1, yielding a distance of $\le 1$kpc; however, they pointed out that Mv can be equal to -8 based on t$_3$ estimated as 34~d, where t$_3$ is the number of days in which a nova wanes
by 3 magnitudes after maximum. This value yields a distance of 1.4\,kpc, which is within the uncertainty of the first estimate from \citet{2008A&A...487..223C}.

If one assumes the distance quoted by \citet{2010ApJ...724..480H}, the mean dust-forming 
region velocity is only $\sim$220\kms. This velocity is the velocity measured in the plane
of the sky, in the direction of the major axis. Owing to the high aspect ratio, a high 
inclination is expected, and this measurement should be considered as close to the 
{\it upper limit} of the velocity field, and one may consider whether such a slowly 
expanding material may imprint its signature in spectroscopic observations. 

\citet{2010PASJ...62L...5S} reported on spectroscopic information of great interest 
to this problem. They observed absorption in the NaI lines detected between -649\kms
to -885\kms with the strongest components at  -750 and -800\kms. One may assume that
these absorption features provide a good approximation of the dust velocity field because they
probe low-excitation cool regions. \citet{2008MNRAS.391.1874D} detected many lines
from these low-ionization species (NaI and MgI) at the earliest stages of the event
(i.e. at the onset of the dust-formation phase).
The impressive narrowness of the features ($\sim$15\kms) can be interpreted as originating from a cool medium. The measured FWHM
can be considered as an upper limit, accounting for the instrumental FWHM of
$\sim$5\kms. Hence, accounting for some turbulence and velocity dispersion of
the expanding material in the FWHM broadening, the thermal broadening may 
originate from material cooler than 1500\,K. 
The \citet{2010PASJ...62L...5S} observations and the 2009 NACO/VISIR 
images were obtained almost at the same epoch in mid 2009. The slower absorbing material in the line of sight was moving 
in 2009 at a velocity of about 650\kms. \citet{2012arXiv1203.6725N} reported on much lower velocities, 350$\pm$160\kms, measured from blue-shifted absorption lines of OI and
SiII. These low velocities were measured around light-curve maximum and increased up to $\sim$500\kms\ just before the dust-formation phase. Using the expansion of VISIR of 0.39$\pm$0.03\,mas/d in the major
axis direction and the slowest velocities of the cool material of 350\kms, one still derives a distance estimate close
to 1.0kpc. This value is a lower estimate, 
since the velocities in the direction of the major axis of the bipolar 
nebula are expected to be higher, while the spectroscopically detected absorption components 
are seen in the line-of-sight direction. 
Lacking an accurate estimate of the inclination, we are left with this
estimate which is potentially affected by larger errors.

To conclude, despite the significant limitations and errors that may 
affect the distance estimate of 1.6kpc$\pm$0.4 provided by \citet{2008A&A...487..223C},
we are reasonably confident that the distance of V1280\,Sco is much 
larger than the distance of 0.6kpc proposed by \citet{2010ApJ...724..480H}.




\subsection{Origin of the bipolarity}

A growing body of observational evidence clearly suggests that many nova ejectae are highly bipolar in 
the first days and months after the outburst. The correlation between speed class and ejection 
velocity, meaning that the faster the expansion speed, the less shaped the remnant 
\citep{2002AIPC..637..497B, 2000AJ....120.2007D}, is an argument for supporting  
the significant effect of the common envelope phase on the shaping.
The ejecta from a slow nova would feel the influence of the secondary much longer, leading
to a strong common envelope interaction. These effects, involving frictional deposition of
energy and angular momentum from the secondary are described by 
\citet{2011ApJ...743..157K,1998MNRAS.296..943P,1997MNRAS.284..137L, 1990ApJ...356..250L}. 
The very slow nova \object{HR\,Del} \citep[t$_3$=230d]{2009AJ....138.1541M,2003MNRAS.344.1219H, 1983ApJ...273..647S} 
or \object{V723\,Cas} \citep[t$_3$=180d]{2009AJ....138.1090L, 2003AJ....126.1981E} are striking examples of this phenomenon.

\object{V1280\,Sco} appears to be at odds with the t$_3$ / axial ratio relationship shown for instance
in Fig.6 of \citet{2002AIPC..637..497B}. The axial ratio of \object{V1280\,Sco} is high, surely higher than 1.5. Such
high axial ratios are exhibited by the slowest novae in the diagram, with t$_3$$>$100d.
\citet{2010ApJ...724..480H} estimated t$_3$ = 34 days, yet cautioned that the fidelity of this determination is weak owing to the complexity of the light curve, which is affected by the 
dust and the secondary maximum. We cannot exclude that t$_3$ might have been shorter, although
given the distance estimate of 1.6kpc that is favored in our study, t$_3$ = 34\,d fits the 
maximum magnitude-rate of decline relation (MMRD) relatively well \citep{2000AJ....120.2007D}.

The late SINFONI images suggest that the dust resides only in the polar caps, and that the equatorial
regions seem devoid of any dust in contrast to the bipolar helium nova \object{V445\,Pup} for
instance, which exhibits a large and dense dusty waist \citep{2009ApJ...706..738W}. The VISIR
last images confirm this since the two lobes appear to be clearly detached from each other in the 
deconvolved images without any sign of emission close to the equatorial plane (Fig.\ref{fig:VISIR3}). \citet{2008A&A...487..223C} reported some evidence for a
significant departure from spherical symmetries in the latest VLTI data (t=145d) from a baseline at more 
than 60 degree from the major axis. A re-inspection of the data shows that contrary to what was suggested
by the authors, the measurements are not affected by any instrumental problem. An aspect ratio as high as 1.3
was probably observed implying that the shaping of the bipolar nebula was already well advanced at this time. 

Two questions can be related to these observations. 
How can this polar ejection be the consequence of a common-envelope phase and what was the 
time scale of the event? With a different time and spatial scale context, the extreme bipolar
planetary nebulae \object{Mz3} and \object{M2-9} can provide enlightening comparisons \citep{2011apn5.confE.305E}. Most of the dust 
in \object{Mz3} and \object{M2-9} was found in the bipolar lobes \citep{2011A&A...527A.105L, 2007A&A...473L..29C, 2005AJ....129..969S}
and recent observations suggest that the time scale of the formation of these nebulae must have been very short,
i.e. a few years at most, in a context of a much slower expected evolution of at least several centuries. 
Regarding \object{M2-9}, it is known that the core harbors a long-period binary \citep[P$\sim$90yr]{2011A&A...529A..43C}.
For these events, the small disks at the core of the systems and the dense lobes favor an intrinsically polar
ejection. This hypothesis cannot be ruled out with the formation of the caps of \object{V1280\,Sco}.

Another scenario implies an intrinsically a-spherical ejection that focuses the mass ejection toward the poles, in which a large-scale structured magnetic field may be the shaping key agent. The recent outburst of the recurrent nova \object{T\,Pyx} appeared to be highly bipolar \citep{2011A&A...534L..11C}. \object{T\,Pyx}  shows a very stable photometric wave that may betray the existence of a magnetic field \citep{1998PASP..110..380P}. The presence of a strong magnetic field of about a few MG might have had a non-negligible effect on the eruption, which is also highly suspected for \object{V1500\,Cyg} stars \citep{2011ATel.3782....1O}.

\section{Conclusion}

We monitored the evolution of the circumstellar material around the classical nova \object{V1280\,Sco}
using high angular resolution infrared observations with the VLT. 
We resolved it for the first time with a single telescope with the NACO/VLT instrument in the near-infrared and VISIR/VLT in the mid-infrared in 2009, two years after the outburst.

These observations revealed a bipolar-shaped nebula around the nova, with dust
present mostly in the lobes of the nebula. The high aspect ratio of the nebula suggests that it is seen at high inclination.

More similar VISIR observations were taken in 2010 and 2011, together with IFU observations with SINFONI/VLT in 2011. This allowed us to observe the expansion of the dusty nebula around V1280 Sco with
a rate of 0.39$\pm$0,03 milliarcsec per day along the major axis of the nebula. Assuming that
the dust we observed with these infrared observations expands at the same speed as the low-excitation
ejecta detected via spectroscopy, this suggests that the distance to V1280 Sco
is at least 1kpc.

Finally, these observations enabled us to study the temporal evolution of a dusty bipolar source.
Further yearly monitoring of \object{V1280\,Sco} will be easier because the nebula is expanding, and will provide an 
excellent test-case for the study of the envelope dust and gas content.

\begin{acknowledgements}
We thank Lowell Tacconi-Garman, Mario van den
Ancker and Elena Valenti, among others, for their invaluable assistance with the preparation of these observations. Part of this work was carried out within the framework of the European Associated Laboratory "Astrophysics Poland-France". We thank Pierre Cruzal\`ebes for his advice on the MARCS synthetic models.
\end{acknowledgements}

\bibliographystyle{aa}
\bibliography{Bib_v1280Sco} 




\end{document}